# Combined microstructural and magneto optical study of current flow in polycrystalline forms of Nd and Sm Fe-oxypnictides

Fumitake Kametani, A. A. Polyanskii, A. Yamamoto, J. Jiang, E. E. Hellstrom, A. Gurevich and D. C. Larbalestier

*Applied Superconductivity Center, National High Magnetic Field Laboratory, Florida State University, Tallahassee FL 32310, USA*

Z. A. Ren, J. Yang, X. L. Dong, W. Lu and Z. X. Zhao

*National Laboratory for Superconductivity, Institute of Physics and Beijing National Laboratory for Condensed Matter Physics, Chinese Academy of Science, P.O. Box 603, Beijing 100190, P.R. China*

**Abstract**

   In order to understand why the *inter-* and *intra*-granular current densities of polycrystalline superconducting oxypnictides differ by three orders of magnitude, we have conducted combined magneto-optical and microstructural examinations of representative randomly oriented polycrystalline Nd and Sm single-layer oxypnictides. Magneto optical images show that the highest $J_c$ values are observed within single grains oriented with their c axes perpendicular to the observation plane, implying that the *intragranular* current is anisotropic. The much lower *intergranular* $J_c$ is at least partially due to many extrinsic factors, because cracks and a ubiquitous wetting As-Fe phase are found at many grain boundaries. However, some grain boundaries are structurally clean under high resolution transmission electron microscopy examination. Because the whole-sample global $J_c(5K)$ values of the two samples examined are 1000-4000 A/cm$^2$, some 10-40 times that found in random, polycrystalline $YBa_2Cu_3O_{7-x}$, it appears that the dominant obstruction to intergranular current flow of many present samples is extrinsic, though some intrinsic limitation of current flow across grain boundaries cannot yet be ruled out.

**Introduction**

   The discovery of superconductivity in the LaFeAsO$_{1-x}$F$_x$ compound [1] has been followed by rapid exploration of many aspects of the superconducting behaviour of the broad class of rare earth iron oxypnictides [2-17] whose transition temperature $T_c$ can reach above 40 K when La is replaced by Ce [5] and above 50 K when the rare earth is Pr, Nd, Sm and Gd [7-11]. Hunte et al. reported that even the La Fe-oxypnictide with Tc ~ 26 K exhibits a very high upper critical field $H_{c2}$ of ~65 T [6] while $H_{c2}$ over 200 T was deduced for the Sm and Nd Fe-oxypnictides [17], strongly suggesting a large high field domain for the Fe-oxypnictides. Foreseeing practical applications, there has been immediate interest in the critical current density too. But all polycrystalline samples of La, Sm and Nd Fe-oxypnictides [12-16] examined to date show signs of less than full grain-to-grain connectivity, raising the same concern of depression of superconducting order parameter at grain boundaries that has so greatly complicated applications of the cuprates [18]. Grain boundary order parameter suppression is fundamental detriment to applications since it means that a randomly aligned grain structure will not pass the full current that can be sustained by intragrain vortex pinning, thus reducing the global or whole-sample current density below that circulating in the grains. In cuprates this depression is very significant causing $J_{cgb}$, the current crossing the grain boundary, depressed exponentially ($J_{cgb} \sim J_c\,e^{-(\theta/\theta_c)}$) as the misorientation $\theta$ rises beyond a critical angle $\theta_c$, where $\theta_c$ is about 3-5° for most cuprate grain boundaries [19][20].

   In a recent study of the magnetization of bulk and powdered samples of polycrystalline La Fe-oxypnictide by Yamamoto *et al.*, very low global current was deduced [12] to flow, leaving open the possibility of an intrinsic granularity similar or even worse than in the cuprates. However this conclusion could not be tested explicitly since the smallest powder size evaluated was ~50μm, several times the grain size. Subsequent study of Sm-[13] and Nd-oxypnictide [14][16] polycrystalline bulks also uncovered evidence for reduced connectivity of polycrystalline sample forms. Our own follow-on study [15] of polycrystalline Sm and Nd oxypnictides showed considerable enhancement of the hysteretic magnetization compared to La-oxypnictide [12]. From sample-size dependent measurements of the magnetization and whole-sample, magneto optical images, we deduced that a significant global current was flowing. However, the intergranular and intragranular current densities had distinctively different temperature dependences and differed in magnitude by a factor of 1000. We also observed that the *intergranular* current density (global $J_c$) of the Sm sample (~ 4000 A/cm$^2$ at 4.2 K) was almost twice higher than that of the Nd sample (~2000 A/cm$^2$) whereas the *intragranular* current density (local $J_c$) was quite similar [15]. In this follow on study, we provide a more detailed and more local correlation between current flow and the microstructure so as to address in greater detail the causes of granularity in the rare earth Fe-oxypnictides.

**Experimental details**

   The polycrystalline SmFeAsO$_{0.85}$ and NdFeAsO$_{0.94}$F$_{0.06}$ bulk samples were synthesized by solid state reaction under high pressure. SmAs (or NdAs) pre-sintered powder and Fe, Fe$_2$O$_3$, and FeF$_2$ powders were mixed together according to the nominal stoichiometric ratio, then ground thoroughly and pressed into small pellets, which were sealed in boron nitride crucibles and sintered under a pressure of 6 GPa at 1250°C for 2 hours [8][10]. This synthesis produces sharp resistive and magnetic $T_c$ transitions, even though the microstructure is far from single phase [15].

MO imaging with a 5 um thick Bi-doped iron-garnet indicator film was used to observe the normal field component B$_z$ produced by magnetization currents induced by applying external fields up to 120 mT perpendicular to the imaging surface [21][22]. Samples were imaged in various states, but the principal one used was that of zero field cooling (ZFC) to the superconducting state, application of fields up to 120 mT, then removal of the field to zero. Such a procedure induces currents to flow throughout the whole sample and allows direct observation of the uniformity of the currents flowing in the sample.

Back scattered electron (BSE) imaging and orientation imaging microscopy (OIM) using electron back scattering diffraction (EBSD) were carried out on well-polished sample surfaces in two scanning electron microscopes (Carl Zeiss 1540 EsB or XB). Inverse pole figure maps were obtained by OIM, in order to highlight the principal (001), (110) and (100) planes intersecting the surface.

Thin lamellae ~10 x 20 um in size were prepared with the focused ion beam tool of the 1540EsB for subsequent transmission electron microscope (TEM) and high resolution TEM (HREM) observation in a JEOL 2011.

**Results and discussion**

Fig 1 shows whole-sample BSE and MO images of the Sm and Nd Fe-oxypnictide samples. Both samples are multi-phase, consisting of the RE Fe-oxypnictide phase (intermediate gray contrast), a glassy Fe-As phase (dark contrast) and Sm$_2$O$_3$ or Nd$_2$O$_3$ (white contrast), as seen in Fig.1 (a) and (c). The area fraction of the Fe-oxypnictide phase calculated with the ImageJ software is ~ 80 % in both samples. The impurity phases are distributed more finely and uniformly in the Sm than in the Nd sample, where the 2$^{nd}$ phase is more inhomogeneous and on a much larger scale of 30-200 um. MO images of the residual magnetic flux of the remnant state in the Sm sample (b) and the Nd sample (d) were taken after zero field cooling (ZFC) to 6 K and applying 120 mT in order to induce whole sample current flow. The bright spots correspond to regions of strongly trapped flux produced by locally high $J_c$ regions. The MO images show that flux penetrated into the center of the Nd sample under an external field of ~10 mT at 6 K while flux reached the center of the Sm sample at the higher external field of ~15 mT, indicating a smaller global circulating current in the Nd sample. The white rectangles in Fig.1 are the areas where we correlate the MO images and the microstructure in detail in Fig.2 and 3.

Typical high $J_c$ bright spots of the Sm sample seen in Fig.1 (b) are black-circled in Fig.2 (a). We should first note that the straight line contrasts visible in the MO images are due to scratches on the MO indicator film and irrelevant to further discussion. Fig.2 (b) shows the inverse pole figure map of the grain orientations in exactly the same region. Several points are clear from this local comparison of MO and OIM images. One is that the grain orientation is essentially random. A second is that the grains are plate-shaped, with an average grain size of ~14 × 6 μm with an aspect ratio of ~0.4 calculated within the OIM scanning area of 105000 μm$^2$ total (not all of which is shown in the figure). Noise on the grain map corresponds to impurity phases such as Fe-As and $Sm_2O_3$. It is clear that most of the bright spots correspond to individual grains of intermediate to large size. Comparing Fig.2(a) and 2(b) where typical high $J_c$ spots A ~ E are marked also suggests that the grains with colours close to red are more likely (i.e. those with grain normal close to [001]) to be high $J_c$ spots, indicating that the strongest MO signals tend to come from the currents circulating on the ab-plane. Some of the bright MO spots also come from intermediate size grains with no preferred crystal orientation, which may imply that grain connectivity in these spots is better than other lower $J_c$ regions, although unfortunately the resolution of the MO images is not quite high enough to show how much current crosses grain boundaries.

In Fig.3, typical high $J_c$ spots in the Nd sample taken from Fig.1 (d) are black-circled in Fig.3 (a) and compared to the inverse pole figure map on exactly the same region in Fig.3 (b). Like the Sm sample, the grain orientation is essentially random. The average grain size of ~7 × 2.8 μm is about half that of the Sm sample but the aspect ratio is also ~0.4. Comparing Fig.3 (a) with (b), the left bright spot comes from the circled large grain whose crystal orientation is shown in the orientation box. The right circle contains two distinct bright spots from two adjacent grains whose orientations are both near [001] and are colored red and pink.

The strong correlation between the microstructure and the high $J_c$ regions in the MO images does suggest that highest density current flows locally within individual grains of both Sm and Nd samples and also that high $J_c$ regions are found preferentially in grains with plane normal close to [001], which also suggests that high $J_c$ occurs for currents flowing on ab-planes, consistent with some superconducting anisotropy. The inverse pole figure maps of Fig.2 (b) and 3 (b) also show clearly that both samples are random polycrystals, meaning that global currents must flow across high angle boundaries.

In spite of the multi-phase microstructure, clean grain boundaries do exist. Fig.4 shows a TEM image of a typical, clean grain boundary in the Sm sample. The image has sharp contrast which rules out any wetting amorphous or impurity phase at the grain boundary. The inset of Fig.4 shows a HREM image of the same grain boundary, in which the sample was tilted so that the grain boundary was almost parallel to the incident electron beam. The lattice fringes of the upper and lower grains impinge at the grain boundary without any diffuse contrast provided by any thin wetting amorphous layer.

However, there are still many non-superconducting obstructions at grain boundaries as clearly seen in the BSE images of Fig.5. Although connected clean grain boundaries are seen in both (a) the Sm and (b) the Nd sample in Fig.5, the Fe-As glassy phase lying between grains and cracks isolate individual grains, limiting the current paths. According to the estimation from Fig.5 (a) and (b) by ImageJ, the length fraction of clean grain boundaries is strongly suppressed down to only ~ 25 % in both Sm and Nd sample, because of this amorphous Fe-As phase, cracks and $Sm_2O_3$ or $Nd_2O_3$.

The two TEM images of Fig.6 (a) and (b) show typical structures of obstructed grain boundaries in the Nd sample. In Fig.6 (a), a current-obstructing crack can be seen at a large angle grain boundary. However, this grain boundary is well-connected at the right side of the same image, at least showing how local the transition from extrinsic limitation of $J_c$ across the grain boundary may be. As shown in Fig.6 (b), while most grain boundaries show solid contrast indicating that they are structurally well-connected, the dark contrast in the BSE image Fig.5 (b) suggests a grain boundary wetted by amorphous phase, providing a second reason for extrinsic obstruction of current at grain boundaries as also indicated in Fig.6 (b). There is also an impurity phase at the GB junction.

The glassy Fe-As phase and $Sm_2O_3$ or $Nd_2O_3$ impurities lying between Fe-oxypnictide grains significantly reduce the current paths in the Sm and Nd samples. The macroscopic inhomogeneity on the scale of several hundreds μm (see Fig.1) substantially disturbs the bulk current over the whole Nd sample as we found in the MO images [15]. In addition, percolation of the supercurrent through a minority of good *intergranular* connections will be forced by the cracks and wetting amorphous phase found at grain boundaries, a state reminiscent of $MgB_2$ where MgO insulating layers at grain boundaries seriously suppress the *intergranular* current [23][24]. In the case of $Bi_2Sr_2Ca_2Cu_3O_x$ textured polycrystalline tapes that are also multi-phase, there is a clear correlation between phase purity and the whole-sample $J_c$, which can suddenly increase by a factor up to 10 times when the volume fraction of the superconducting phase exceeds a certain threshold [25][26]. Based on the differences of microstructure and MO response observed here for the Nd and Sm samples, we suppose that the difference of ~2 between the whole-sample $J_c$ of the Sm and Nd samples results from differences in the extrinsic factors (macroscopic phase inhomogeneity, grain boundary cracks, and wetting amorphous Fe-As phase at grain boundaries) rather than intrinsic property variation.

At this stage of Fe-oxypnictide studies, rather few reports of the phase state and its influence on $J_c$ have yet been made, making firm conclusions hard to draw. Prozorov et al. carried out MO imaging on a $NdFeAsO_{0.9}F_{0.1}$ bulk in which remnant field was trapped only in individual grains, showing strong granularity too [14]. Moore et al also showed that only small current flows over macroscopic dimensions in a $NdFeAsO_{0.85}$ bulk. They too found a wetting phase around the Nd oxypnictide grains [16]. Senatore *et al*. reported impurity phases in their $SmO_{0.85}F_{0.15}FeAs$ sample which also showed a significant sign of weak-link behaviour [13]. In fact a reasonable conclusion is that all polycrystalline RE Fe-oxypnictide samples reported so far are multi-phase. In this important respect, therefore, we believe that the samples described here are fully representative of present polycrystalline materials.

Even with the *intergranular* $J_c$ limitation by multiple extrinsic factors, the global $J_c$ is at least 10 times higher than that in a random polycrystalline ReBCO [27][28], where values of $J_c(4K) \sim 100$ A/cm$^2$ are found in single-phase samples with clean grain boundaries. This comparison suggests a much less strong intrinsic weak-link effect at grain boundaries in the oxypnictides than in the cuprates. In these samples, $Sm_2O_3$ and $Nd_2O_3$ are completely insulating and serious blocks to *intergranular* current flow. Nor can we expect large current flow across the glassy Fe-As phase even though Yamamoto *et al.* found an SNS component to the *intergranular* flow that is consistent with SNS coupling across this phase. Considering that only a few of the grain boundaries are cleanly coupled without extensive secondary phase of the type seen in Figs. 5 and 6, it is reasonable to think that the global $J_c$ of the Sm and Nd samples is potentially much higher than what we have reported [15]. In order to better understand the intrinsic weak-link effects at grain boundaries, we need to make bulk samples of much higher phase purity and to examine current dissipation on single grain boundaries [29] of defined misorientation.

**Conclusion**

We have investigated the causes of two distinct scale of current and different intergranular current density observed in the polycrystalline Sm and Nd Fe-oxypnictides. We find that impurity phases extrinsically limit the intergranular current on the macro scale. High-density current flows locally within individual grains, preferentially circulating on ab-planes. However, clean grain boundaries without any wetting amorphous phase were found too. The difference of global $J_c$ between the Nd and Sm samples appears to result from macroscopic inhomogeneity, and cracks and wetting amorphous at grain boundaries. Considering their random polycrystalline form, we conclude that extrinsic limitation of current is still dominant in these Sm and Nd Fe-oxypnictides and that the *intergranular* intrinsic limitation is less severe than in the cuprates.
.

**Acknowledgments**

Work at the NHMFL was supported by IHRP 227000-520-003597-5063 under NSF Cooperative Agreement DMR-0084173, by the State of Florida, by the DOE, by the NSF Focused Research Group on Magnesium Diboride (FRG) DMR-0514592 and by AFOSR under grant FA9550-06-1-0474.

**Figure captions**

Fig.1 Back scatter electron (BSE) and magneto optical (MO) images of the Sm sample (a) (b), and on the Nd sample (c) (d), respectively. MO imaging was done after ZFC to 6 K, then applying 120 mT and then reducing the field to 0 mT.

Fig.2 (a) Typical high $J_c$ bright spots in the MO image of the Sm sample taken from Fig.1 (b). The straight line contrasts visible in the MO images are due to scratches on the MO indicator film and irrelevant to further discussion. (b) Inverse pole figure map of the exact same region. Black-circled areas correspond to the high $J_c$ spots in (a).

Fig.3 (a) Typical high $J_c$ bright spots of the MO image on the Nd sample taken from Fig.1 (d). (b) Inverse pole figure map of the exact same region. Black-circled areas correspond to the high $J_c$ spots in (a).

Fig.4 TEM image showing a clean grain boundary in the Sm sample. The inset of HREM image of the same grain boundary proves no thin wetting amorphous on GB.

Fig.5 BSE image of the (a) Sm and (b) Nd sample at high magnification. Although some grain boundaries are well-connected, others are clearly obstructed by the Fe-As phase (dark contrast), $Sm_2O_3$ or $Nd_2O_3$ (white contrast) and cracks.

Fig.6 TEM images showing grain boundaries in the Nd sample obstructed by (a) cracks and (b) the wetting amorphous phase and $Nd_2O_3$.

Figure 1

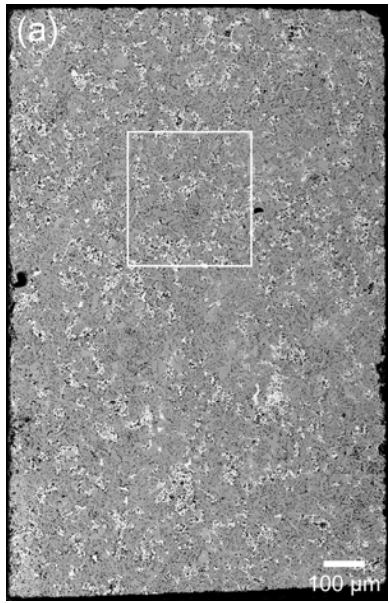 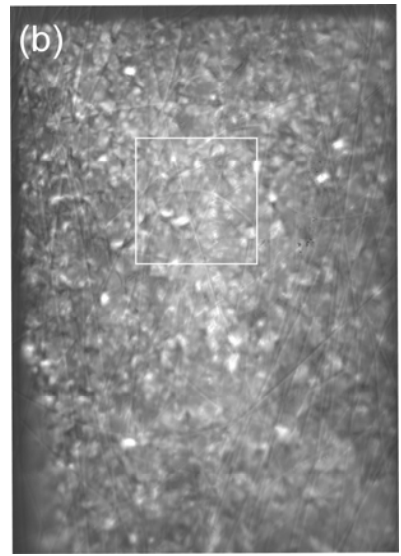

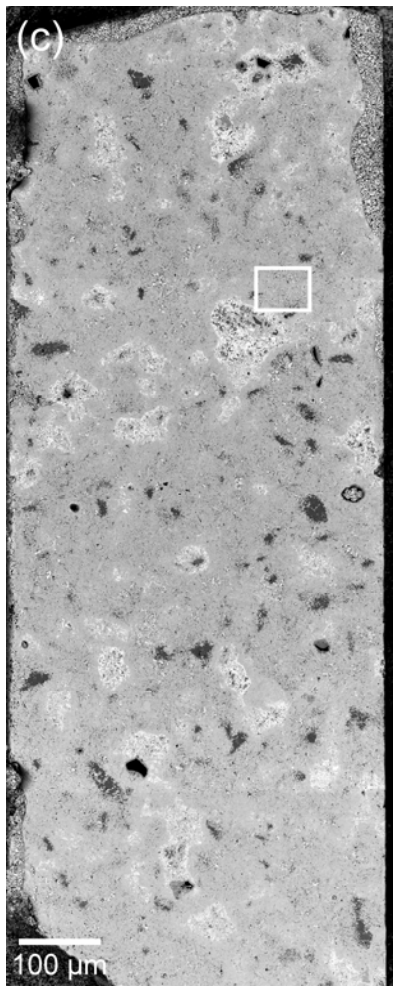 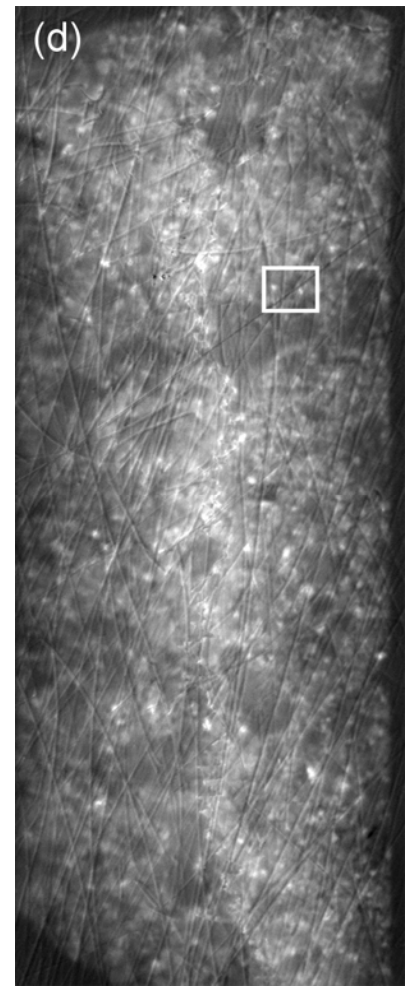

Figure 2

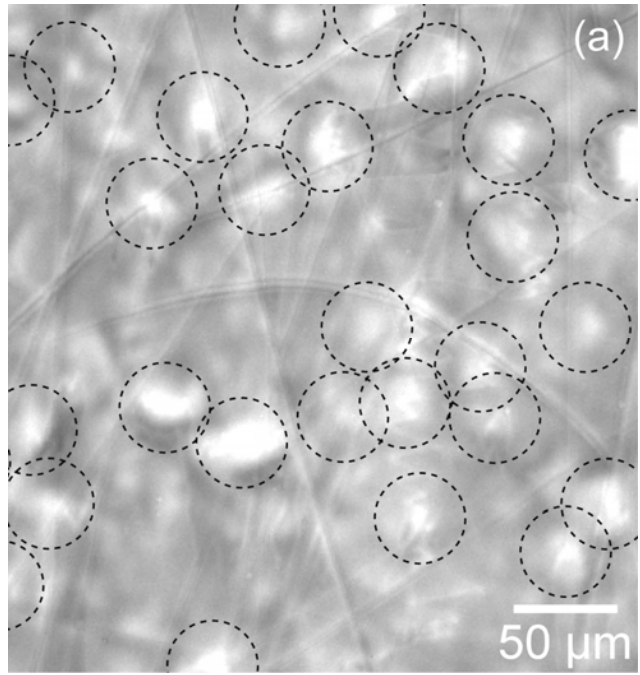

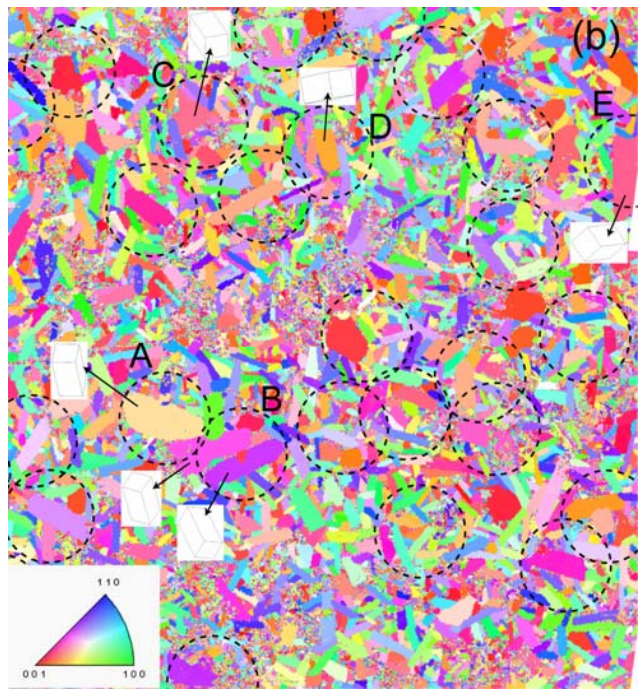

Figure 3

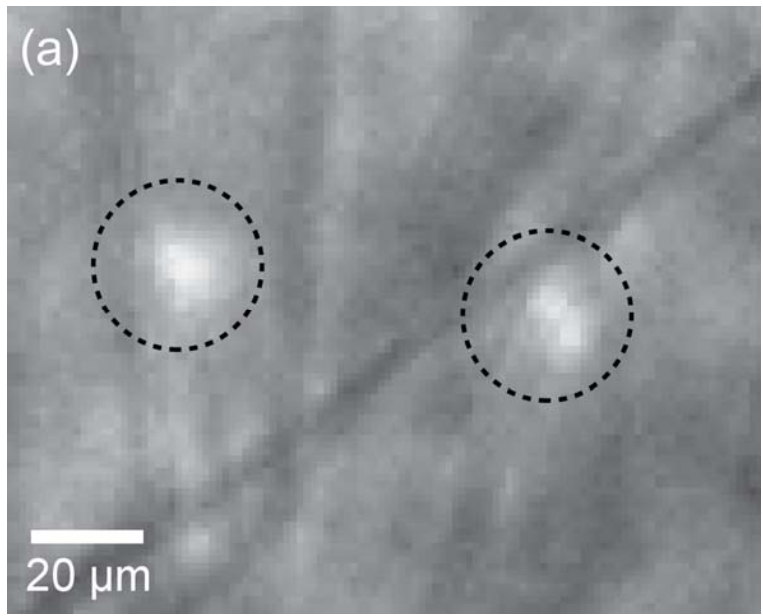

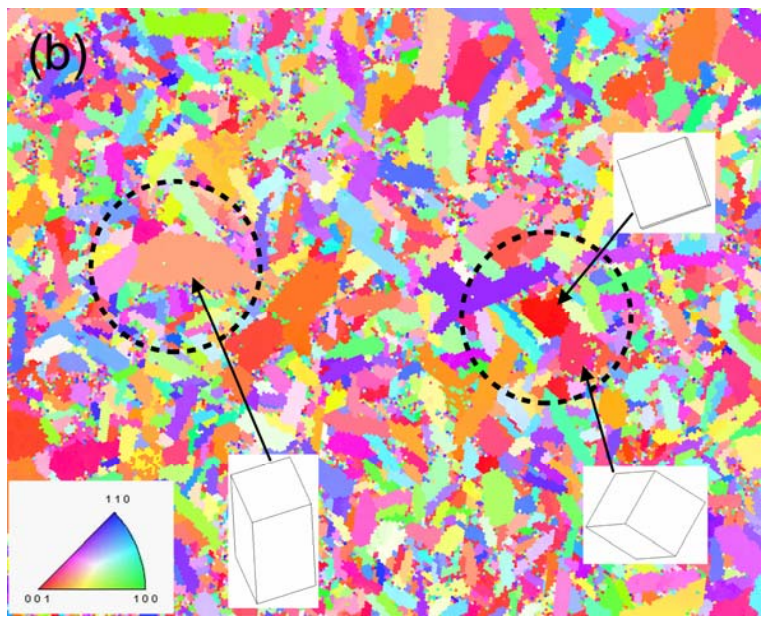

Figure 4

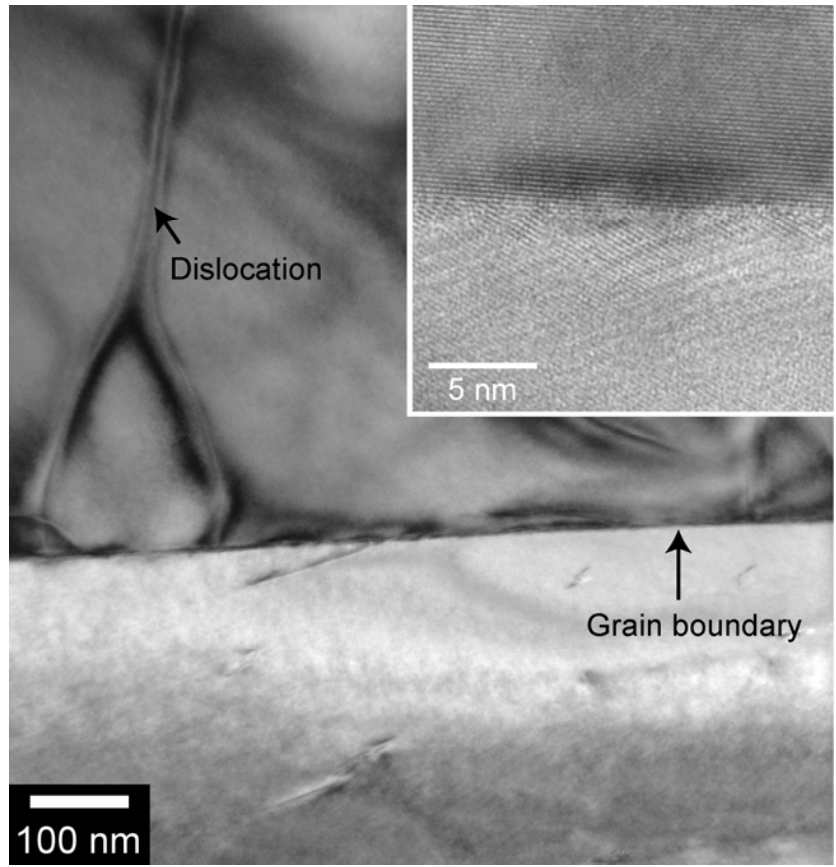

Figure 5

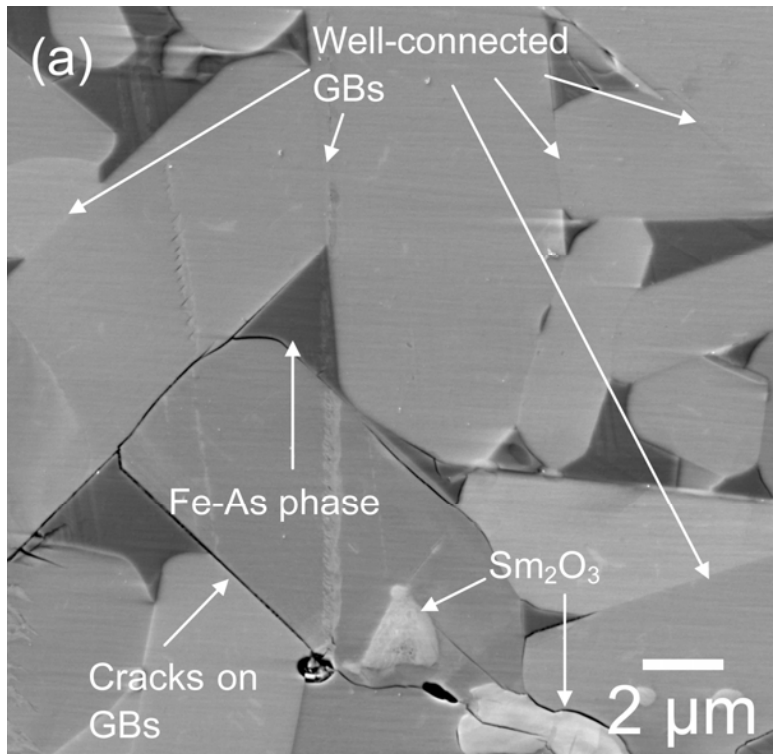

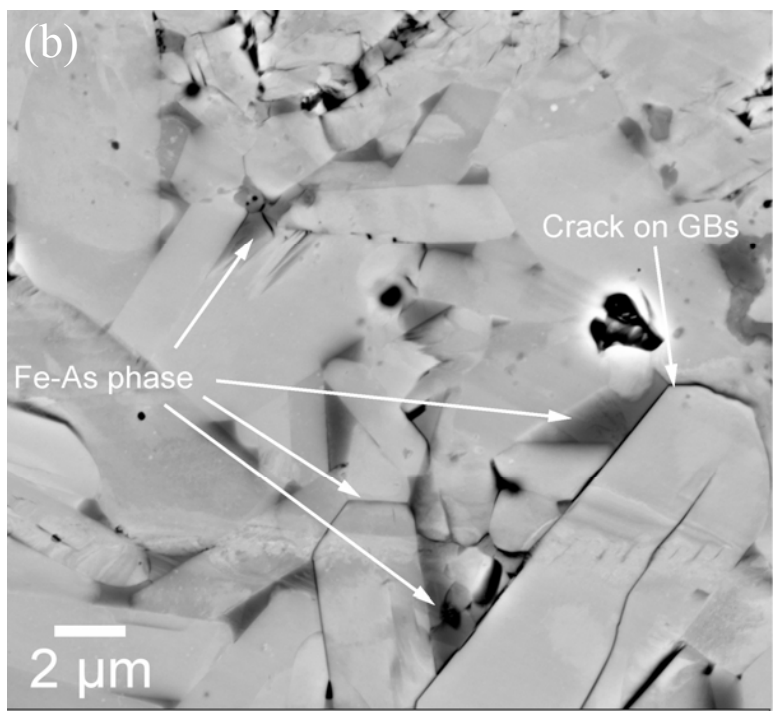

Figure6

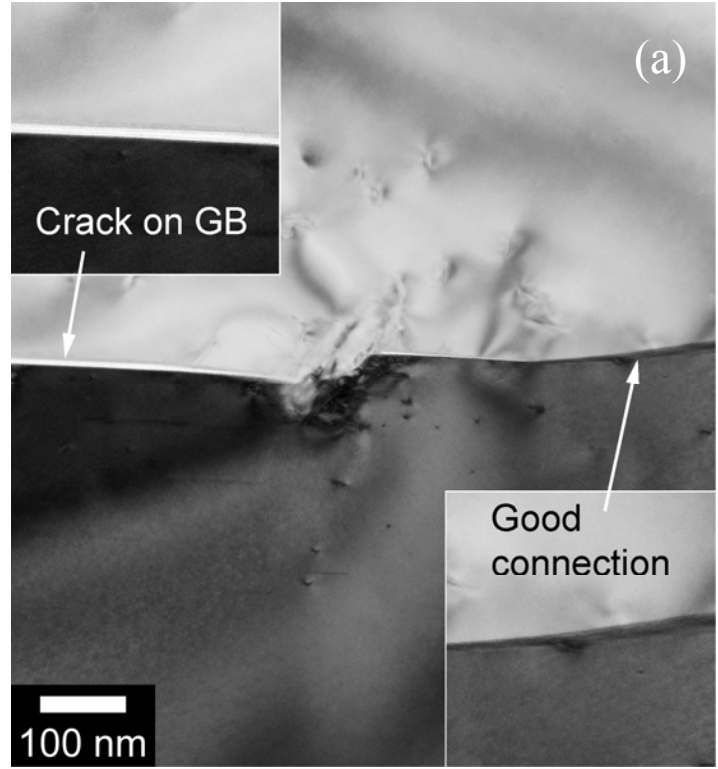

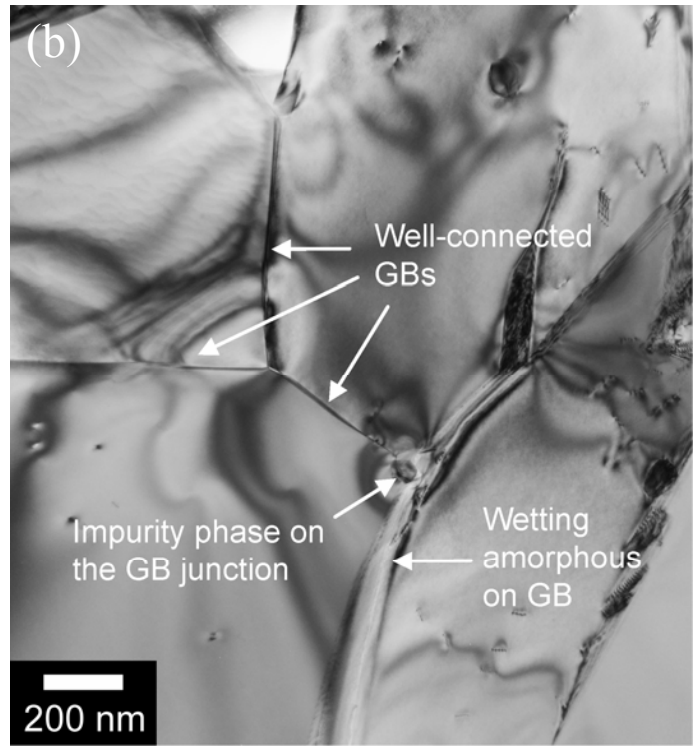